\documentclass{article}
\usepackage{romp32}

\input{Qcircuit}

\renewcommand{\Qcircuit}[1][0em]{\xymatrix @*[o] @*=<#1>}
\newcommand{\node}[2][]{{\begin{array}{c} \ _{#1}\  \\ {#2} \\ \
\end{array}}\drop\frm{o} }
\newcommand{\link}[2]{\ar @{-} [#1,#2]}
\newcommand{\halflink}[2]{\ar @{.} [#1,#2]}

\title{Cluster-state quantum computation}
\author{
  Michael A. Nielsen \\
  School of Physical Sciences, The University of Queensland, \\
  Brisbane, Queensland 4072, Australia \\
  nielsen@physics.uq.edu.au and www.qinfo.org/people/nielsen/}

\begin{document}

\maketitle
\begin{abstract}
  This article is a short introduction to and review of the
  \emph{cluster-state} model of quantum computation, in which coherent
  quantum information processing is accomplished via a sequence of
  single-qubit measurements applied to a fixed quantum state known as
  a \emph{cluster state}.  We also discuss a few novel properties of
  the model, including a proof that the cluster state cannot occur as
  the exact ground state of any naturally occurring physical system,
  and a proof that measurements on any quantum state which is linearly
  prepared in one dimension can be efficiently simulated on a
  classical computer, and thus are not candidates for use as a
  substrate for quantum computation.
\end{abstract}


\section{Introduction}

Every child who has played computer games understands intuitively that
one physical system can simulate another.  Despite the simplicity of
this idea, it is only recently that researchers have begun to develop
a deep understanding of simulation from the point of view of basic
physics.

Perhaps the fundamental question to be answered about simulation is:
``when can one physical system be used to efficiently simulate
another?''  Recent work on this question has been motivated by the
blossoming of interest in quantum computers~\cite{Nielsen00a}, as-yet
hypothetical devices which, it is hoped, can be used to efficiently
simulate any other physical system.  Thus, by asking what physical
resources are \emph{universal} for quantum computation, i.e., can be
used to build a quantum computer, we are asking a general question
about what physical resources are sufficient to efficiently simulate
any other physical system.

The purpose of the present paper is to review recent work on
\emph{measurement-based} quantum computation, i.e., models for quantum
computation having the remarkable property that all the basic
dynamical operations are non-unitary quantum measurements, yet they
can still be used to simulate arbitrary quantum dynamics, including
unitary dynamics.  Such models of quantum computation thus challenge
the conventional understanding of quantum measurement as a process
that inherently destroys quantum coherence.

We focus on a class of measurement-based models of quantum computation
proposed by Raussendorf and Briegel~\cite{Raussendorf01a}, the
so-called \emph{cluster-state} model, or \emph{one-way quantum
  computer}.  The cluster-state model has a remarkably rich structure
that is not fully understood, but which differs substantially from the
conventional unitary model of quantum computing.  These differences
have led to new insights into quantum computational
complexity~\cite{Raussendorf02b}, and to dramatic simplifications in
experimental proposals for quantum
computation~\cite{Nielsen04b,Browne04a,Walther05a}.

An alternate approach to measurement-based quantum computation has
been proposed in~\cite{Nielsen03b}, and developed
in~\cite{Leung01c,Leung03a}.  We will not explore this other approach
here, but note that connections between the two approaches have been
developed in~\cite{Aliferis04a,Childs04a,Jorrand04a}.  More generally,
the literature on measurement-based quantum computation has grown
rapidly, and we cannot do a thorough survey of all developments here;
see, for
example,~\cite{Rudolph05a,Aliferis05a,Danos04a,Mhalla04a,Nielsen04c,Perdrix04a,Verstraete04a,Barrett04a,Hutchinson04a,Raussendorf03a,Raussendorf02b,Raussendorf02a}
and references therein for more information.

The structure of the paper is as follows.  Section~2 presents a short
review of the standard unitary quantum circuits model of quantum
computation.  Section~3 is an elementary introduction to and review of
the basic cluster-state model, explaining the model and how it can be
used to simulate the quantum circuit model of computation.  In
addition to this review function, the paper also contains in Section~4
a discussion of some novel results about two open problems, namely:
(a) can the cluster state arise as the ground state of some reasonable
physical system; and (b) what quantum states, when used as a
substrate, can be efficiently simulated on a conventional classical
computer?  For neither problem do we obtain anything like a
comprehensive solution, yet in both cases we obtain through elementary
means results that hint at a beautiful structure yet to be fully
understood.  Section~5 concludes.

It is a pleasure to dedicate this paper to Tony Bracken.  As a
student, I had the good fortune to receive Tony's supervision for an
Honours project on quasidistribution functions in quantum mechanics.
That project combined in a pleasing way fundamental physical questions
with simple yet beautiful and sometimes surprising mathematics.  I
hope the present subject offers something of the same character to
readers.

\section{The quantum circuit model of quantum computation}
\label{sec:standard_qc}

\emph{A priori} there are many ways one might construct a
quantum-mechanical model of computation.  To date, all the physically
plausible models of quantum computation which have been proposed have
turned out to be computationally equivalent.  That is, each of these
apparently different models of computation can efficiently simulate
one another, and thus do not differ in the class of computational
problems they can efficiently solve.

The most widely used model of quantum computation at the present time
is the \emph{quantum circuit} model~\cite{Deutsch89a}, which is a
generalization of the well-known classical circuit model based on
Boolean logical operations such as {\sc and} and {\sc not}.  We
provide a brief review of the quantum circuit model in this section,
referring the reader to Chapter~4 of~\cite{Nielsen00a} for more
details.  Through the remainder of this paper we treat the quantum
circuit model as the standard model of quantum computation.

The main elements of the quantum circuit model are illustrated by the
following example of a quantum circuit:
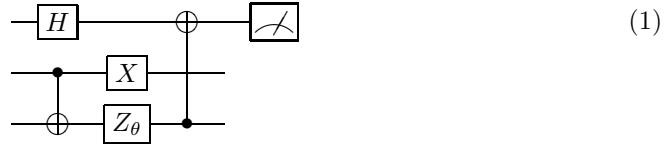
\begin{equation} \label{eq:example-circuit}
\Qcircuit @C=1em @R=0.5em @!R {
& \gate{H} & \qw             & \targ     & \qw & \meter  \\
& \ctrl{1} & \gate{X}        & \qw       & \qw & \\
& \targ    & \gate{Z_\theta} & \ctrl{-2} & \qw &
}
\end{equation}
The horizontal lines are \emph{quantum wires}, representing
\emph{qubits}, abstract quantum mechanical systems with a
two-dimensional state space spanned by the (orthogonal)
\emph{computational basis states} $|0\rangle$ and $|1\rangle$.  The
left-to-right progress of a wire does not represent movement of the
qubit through space, but rather progress through time.  The initial
state of the qubits is usually taken to be some fiducial product
state, such as the all $|0\rangle$ state, $|0\rangle^{\otimes n}$.

The processing of the qubits is done through a sequence of one- and
two-qubit quantum gates.  These quantum gates are unitary operations
transforming the state of one or two qubits.  For example, we have
single-qubit gates like the Hadamard gate,
\begin{equation}
\Qcircuit @C=1em @R=0.5em @!R {
& \gate{H} & \qw} \equiv
 \frac{1}{\sqrt 2} \left[ \begin{array}{cc} 1 & 1 \\ 1 & -1 \end{array} 
 \right],
\end{equation}
where the matrix on the right is the unitary matrix representing the
action of the Hadamard gate with respect to the $|0\rangle, |1\rangle$
basis.  So, for example, the Hadamard gate takes the input $|0\rangle$
and transforms it to $(|0\rangle+|1\rangle)/\sqrt 2$.

Other important single-qubit gates are the rotations about the $X$,
$Y$ and $Z$ axes:
\begin{eqnarray}
\Qcircuit @C=1em @R=0.5em @!R {
& \gate{X_(\theta)} & \qw} & \equiv &
 \exp(-i \theta X/2) \\
\Qcircuit @C=1em @R=0.5em @!R {
& \gate{Y_(\theta)} & \qw} & \equiv &
 \exp(-i \theta Y/2) \\
\Qcircuit @C=1em @R=0.5em @!R {
& \gate{Z_(\theta)} & \qw} & \equiv &
 \exp(-i \theta Z/2),
\end{eqnarray}
where $X, Y$ and $Z$ are abbreviated notations for the usual Pauli
$\sigma_x, \sigma_y$ and $\sigma_z$ matrices.

Perhaps the most commonly-used two-qubit gate is the controlled-{\sc
  not} gate,
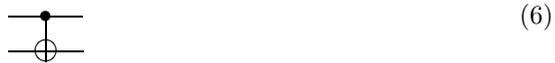
\begin{equation}
\Qcircuit @C=1em @R=0.5em @!R {
 & \ctrl{1} & \qw \\
 & \targ    & \qw
}
\end{equation}
The top qubit is the \emph{control} qubit, while the bottom is the
\emph{target} qubit.  These names are used because the action of the
controlled-{\sc not} in the computational basis is to take
$|x,y\rangle$ to $|x,y\oplus x\rangle$, where $\oplus$ is addition
mod~$2$.  That is, the control qubit remains unchanged, while the
target is flipped if the control is set to $1$, and is otherwise
unchanged.

The controlled-{\sc not} is a special case of more general
controlled-unitary gates:
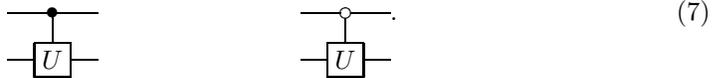
\begin{eqnarray}
\Qcircuit @C=1em @R=0.5em @!R {
 & \ctrl{1} & \qw \\
 & \gate{U} & \qw
} 
& \hspace{2cm} & 
\Qcircuit @C=1em @R=0.5em @!R {
 & \ctrlo{1} & \qw \\
 & \gate{U}  & \qw
}.
\end{eqnarray}
In both cases the first qubit is the control, while the second qubit
is the target.  In the first circuit, $U$ is applied to the target
qubit if the control qubit is set to $1$, while in the second circuit
$U$ is applied to the target qubit if the control qubit is set to $0$.

Beside the controlled-{\sc not}, another specific controlled-unitary
gate that is often useful is the controlled-{\sc phase} gate,
\begin{eqnarray}
\Qcircuit @C=1em @R=1.8em @!R {
 & \control \qw  & \qw \\
 & \control \qw \qwx & \qw
} 
& \equiv & 
\Qcircuit @C=1em @R=0.5em @!R {
 & \ctrl{1} & \qw \\
 & \gate{Z} & \qw
} 
=
\Qcircuit @C=1em @R=0.5em @!R {
 & \gate{Z} & \qw \\
 & \ctrl{-1} & \qw
}
\end{eqnarray}
The action of the controlled-{\sc phase} in the computational basis is
$|x,y\rangle \rightarrow (-1)^{xy} |x,y\rangle$.

Since the individual gates are unitary, the combined effect of the
gates making up a circuit such as~(\ref{eq:example-circuit}) is a
joint unitary transformation on the $n$ input qubits.  Which unitary
operations can be synthesized in this way?  Provided the available
gates include all single-qubit gates, and at least one entangling
two-qubit gate, such as the controlled-{\sc not} or controlled-{\sc
  phase}, it turns out that the gate set is \emph{universal}, meaning
that it can be used to synthesize an arbitrary unitary operation on
the qubits~\cite{Brylinski02a}.

The challenge of quantum computation is to find small quantum circuits
synthesizing desirable unitary operations.  For a generic unitary $U$
on $n$ qubits, the number of gates required to synthesize $U$ scales
exponentially in $n$~\cite{Knill95a}.  Such exponential scaling is
prohibitively expensive, and it is far more desirable to find families
of unitary operations which can be synthesized using a number of gates
that scales polynomially in $n$.

The final step in a quantum circuit is to \emph{read out} the state of
the qubits, or some subset of the qubits.  This is done by measuring
the desired subset of the qubits in their respective computational
bases.  The resulting string of bits ``$0010110 \ldots$'' is the
result of the computation.  This idea is illustrated
in~(\ref{eq:example-circuit}), where the meter terminating the top
wire indicates a measurement in the computational basis.

%
% variations
%
This concludes our basic overview of the quantum circuit model of
computation.  In practice, many straightforward variants of the
quantum circuit model are often used.  These include: (1) allowing the
input state to be any tensor product of single-qubit states,
$|\psi_1\rangle \otimes |\psi_2\rangle \otimes \ldots \otimes
|\psi_n\rangle$; (2) allowing measurements with respect to any
orthonormal single-qubit basis, since this is equivalent to applying a
single-qubit unitary operation followed by a computational basis
measurement; and (3) allowing measurements and feedforward of the
measurement results during the computation, so later actions (e.g.,
quantum gates) may depend on the results of earlier measurement
outcomes.  None of these modifications changes the computational power
of the quantum circuit model, but it does make it more convenient to
work with.

\section{The cluster state model}
\label{sec:csc}

In this section we explain the cluster-state model of quantum
computation, and explain how cluster states can be used to efficiently
simulate quantum circuits.  Our discussion
follows~\cite{Nielsen04c,Nielsen03c}, which are based in turn on the
original paper~\cite{Raussendorf01a}.

\subsection{How a cluster-state computation works}

A cluster-state computation begins with the preparation of a special
entangled many-qubit quantum state, known as a \emph{cluster state},
followed by an adaptive sequence of single-qubit measurements, which
process the cluster, and finally read-out of the computation's result
from the remaining qubits.  We now discuss each of these steps in
detail.

%
% define cluster states
%
The term ``cluster state'' refers not to a single quantum state, but
rather to a family of quantum states.  The idea is that to any graph
$G$ on $n$ vertices we can define an associated $n$-qubit cluster
state, by first associating to each vertex a corresponding qubit, and
then applying a graph-dependent preparation procedure to the qubits,
as described below.  As an example, the following graph represents a
six-qubit cluster state,
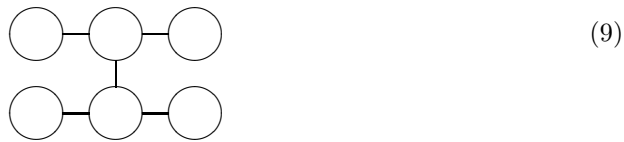
\begin{equation}
\Qcircuit[2em] @R=1em @C=1em {
\node{} & \node{} \link{0}{-1} \link{1}{0} & \node{} \link{0}{-1} \\
\node{} & \node{} \link{0}{-1}             & \node{} \link{0}{-1}
}
\end{equation}
The cluster state associated to the graph may be defined as the result
of applying the following preparation procedure:
\begin{enumerate}
\item Prepare each of the $n$ qubits in the state $|+\rangle \equiv
  (|0\rangle+|1\rangle)/\sqrt 2$.
 
\item Apply controlled-{\sc phase} gates between qubits whose
  corresponding graph vertices are connected.
\end{enumerate}
Note that controlled-{\sc phase} gates commute with one another, so we
do not need to specify the order in which the gates are applied. Also,
although we have described the preparation of the cluster in terms of
applying quantum gates, later in the paper we briefly describe how to
prepare clusters using measurements alone, and so the cluster-state
model may be regarded as a truly measurement-only model of quantum
computation.

%
% nomenclature
%
Note that the states we have called cluster states are sometimes also
known as \emph{graph states}.  Originally, the term ``cluster state''
was introduced by Raussendorf and Briegel~\cite{Raussendorf01b} to
refer to the case where the graph $G$ is a two-dimensional square
lattice.  This was the class of states which they showed
in~\cite{Raussendorf01a} could be used as a substrate for quantum
computation.  The term ``graph state'' originally referred to the
family of states associated with more general graphs $G$.  This
distinction was blurred by the introduction of schemes for quantum
computing based on Raussendorf and Briegel's ideas, but using
different graphs.

I believe it makes most sense to have a single terminology for the
entire class of states, and then to specify in any instance what graph
is being used (e.g. a two-dimensional square lattice with boundary).
I suggest using the term ``cluster state'' for this purpose, and will
follow this terminology throughout this paper.

Once the cluster state is prepared, the next step in the computation
is to perform a sequence of \emph{processing measurements} on the
state.  These measurements satisfy: (1) they are single-qubit
measurements; (2) the choice of measurement basis may depend on the
outcomes of earlier measurements, i.e., feedforward of classical
measurement results is allowed; and (3) measurement results may be
processed by a classical computer to assist in the feedforward, so the
choice of basis may be a complicated function of earlier measurement
results.  Note that for the cluster-state computation to be
\emph{efficient} we must constrain the classical computation to be of
polynomial size.

The output of the cluster-state computation may be defined in two
different ways, both useful.  The first is to regard the computation
as having a \emph{quantum state} as output, namely, the quantum state
of the qubits which remain when the sequence of processing
measurements has terminated.  The second definition is to add a set of
\emph{read-out measurements}, a sequence of single-qubit measurements
applied to the qubits which remain when the processing measurements
are complete.  In this case the output of the computation is a
classical bit string.

A concrete example of these ideas is the following cluster-state
computation:
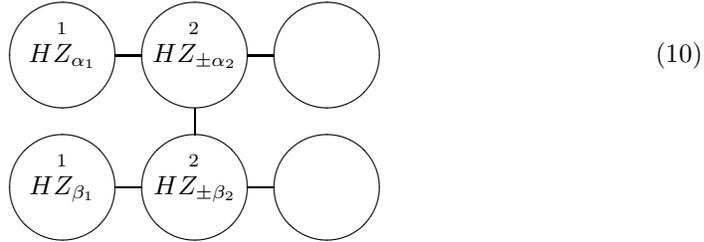
\begin{equation} \label{eq:first-cluster-figure}
\Qcircuit[4em] @R=1em @C=1em {
\node[1]{HZ_{\alpha_1}} & \node[2]{HZ_{\pm \alpha_2}} \link{0}{-1} \link{1}{0} 
    & \node{} \link{0}{-1} \\
\node[1]{HZ_{\beta_1}}  & \node[2]{HZ_{\pm \beta_2}} \link{0}{-1} 
    & \node{} \link{0}{-1}
}
\end{equation}
Labels indicate qubits on which processing measurements occur, while
unlabeled qubits are those which remain as the output of the
computation when the processing measurements are complete.  Note that
the qubits are labeled by a positive integer $n$ and a single-qubit
unitary, which we refer to generically as $U$; here $U = HZ_{\pm
  \alpha_j}, HZ_{\pm \beta_j}$.  The $n$ label indicates the
time-ordering of the processing measurements, with qubits having the
same label capable of being measured in either order, or
simultaneously.  The time order is important, because it determines
which measurement results can be fedforward to control later
measurement bases.  The $U$ label indicates the basis in which the
qubit is measured, denoting a rotation by the unitary $U$, followed by
a computational basis measurement.  Equivalently, a single-qubit
measurement in the basis $\{ U^\dagger |0\rangle, U^\dagger |1\rangle
\}$ is performed.  The $\pm$ notation in $HZ_{\pm \alpha_2}$ and
$HZ_{\pm \beta_2}$ indicates that the choice of sign depends on the
outcomes of earlier measurements, in a manner to be specified
separately.  We'll give an example of how this works later.

\subsection{Simulating quantum circuits in the cluster-state model}

We now explain how quantum circuits can be simulated using a
cluster-state computation.  The key idea underlying the simulation is
a simple circuit identity, sometimes known as one-bit
teleportation~\cite{Zhou00a}:
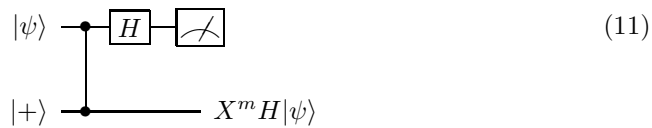
\begin{equation} \label{eq:basic-transport}
\Qcircuit @C=1em @R=1.8em @!R {
 \lstick{|\psi\rangle} & \ctrl{1}     & \gate{H} & \meter \\
 \lstick{|+\rangle}    & \control \qw & \qw      & \rstick{X^m H|\psi\rangle}
 \qw
}
\end{equation}
Here $m$ is the outcome (zero or one) of the computational basis
measurement on the first qubit.  This identity may be verified by
expanding $|\psi\rangle = \alpha|0\rangle+\beta|1\rangle$, so the
state after the controlled-{\sc phase} and Hadamard gates is
$\alpha|++\rangle+\beta|--\rangle$, by the gate definitions given
earlier.  This state may be re-expressed as $(|0\rangle \otimes
H|\psi\rangle + |1\rangle \otimes XH |\psi\rangle)/\sqrt 2$, from
which the result follows.

The identity of~(\ref{eq:basic-transport}) is easily generalized to
the following identity:
\begin{equation} \label{eq:transport}
\Qcircuit @C=1em @R=1.8em @!R {
 \lstick{|\psi\rangle} & \ctrl{1}     & \gate{HZ_\theta} 
 & \meter \\
 \lstick{|+\rangle}    & \control \qw & \qw      
 & \rstick{X^m H Z_\theta |\psi\rangle} \qw
}
\end{equation}
The proof is to note that $Z_\theta$ commutes with the controlled-{\sc
  phase} gate, and thus the output of the circuit is the same as would
have been output from the circuit in
Equation~(\ref{eq:basic-transport}) had $Z_\theta |\psi\rangle$ been
input, instead of $|\psi\rangle$.

The proof of~(\ref{eq:transport}) is elementary, but the result is
nonetheless remarkable. Observe that although the first qubit is
measured, no quantum information is lost, for no matter what the
measurement outcome, the posterior state of the second qubit is
related by a known unitary transformation to the original input,
$|\psi\rangle$.

It is tempting to regard this as unsurprising.  After all, suppose we
replaced the controlled-{\sc phase} gate by a {\sc swap} gate, which
merely interchanges the state of the two qubits.  Then we would not
expect a measurement on the first qubit to destroy any quantum
information, since all the quantum information would have been
transferred from qubit one to qubit two \emph{before} the measurement
on qubit one.

However, this is not what happens, as can be seen from the fact that
by varying the basis in which the first qubit is measured, i.e., by
varying $\theta$, we can vary the unitary transformation effected on
the second qubit, without destroying any quantum information.  This
may be regarded as a generalization of the EPR
effect~\cite{Einstein35a}, and may also be viewed as an instance of a
quantum error-correcting code (see, e.g., Chapter~11
of~\cite{Nielsen00a}, and compare with the two-qubit error-detection
code in~\cite{Knill00a}.)

We can use~(\ref{eq:transport}) to explain how cluster-state
computation can simulate quantum circuits.  We begin by explaining how
to simulate a single-qubit circuit of the form:
\begin{equation} \label{eq:basic-one-qubit-circuit}
\Qcircuit @C=1em @R=1.8em @!R {
 \lstick{|+\rangle} & \gate{HZ_{\alpha_1}} & \gate{HZ_{\alpha_2}} & \qw
}  
\end{equation}
This apparently trivial case contains the most important ideas used in
the general case. Note that we assume the qubit starts in the
$|+\rangle$ state, and that single-qubit gates are of the form
$HZ_\alpha$.  These assumptions are made for convenience, and do not
cause any loss of generality, since it is clear that an arbitrary
single-qubit circuit can be simulated using the ability to simulate
these operations.

The cluster-state computation used to simulate
circuit~(\ref{eq:basic-one-qubit-circuit}) is:
\begin{equation}
\Qcircuit[4em] @R=1em @C=1em {
 \node[1]{HZ_{\alpha_1}} & \node[2]{HZ_{\pm \alpha_2}} \link{0}{-1} &
\node{} \link{0}{-1}
}    
\end{equation}
By definition, this cluster-state computation has an output equal to
the output of the following quantum circuit\footnote{Note that the
  double vertical lines emanating from the meter on the top qubit
  indicate classical feedforward and control of later operations.  We
  use this and similar notations often later in the paper.}:
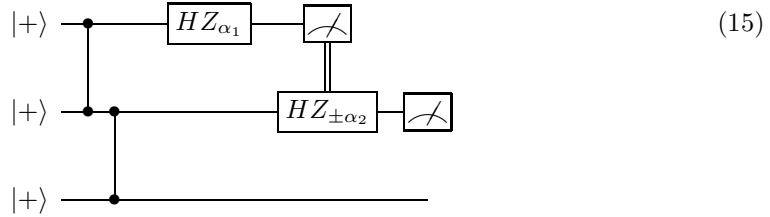
\begin{equation}
\Qcircuit @C=1em @R=1.8em @!R {
 \lstick{|+\rangle} & \ctrl{1}     & \qw          & \qw & 
   \gate{HZ_{\alpha_1}} & \meter \\
 \lstick{|+\rangle} & \control \qw & \ctrl{1}     & \qw &
   \qw                 & \gate{HZ_{\pm \alpha_2}} \cwx & \meter \\
 \lstick{|+\rangle} & \qw          & \control \qw & \qw &
   \qw                  & \qw                          & \qw
}  
\end{equation}
Equivalently, we can delay the operations on the second and third
qubits until \emph{after} the measurement on the first qubit is
complete:
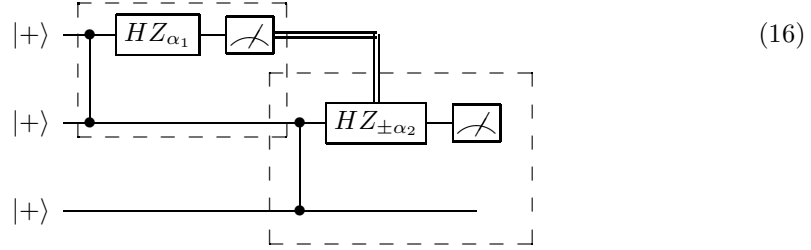
\begin{equation}
\Qcircuit @C=1em @R=1.8em @!R {
 \lstick{|+\rangle} & \ctrl{1}     & \gate{HZ_{\alpha_1}} & \meter & 
\cw          & \cw \cwx[1] \\
 \lstick{|+\rangle} & \control \qw & \qw                  & \qw    &
\ctrl{1}     & \gate{HZ_{\pm \alpha_2}} & \meter \\
 \lstick{|+\rangle} & \qw          & \qw                  & \qw    & 
\control \qw & \qw                      & \qw 
\gategroup{1}{2}{2}{4}{1em}{--}
\gategroup{2}{5}{3}{7}{2.4em}{--}
}  
\end{equation}
To determine the output, observe that the two highlighted boxes are
both of the form of~(\ref{eq:transport}), and thus the output of the
circuit is $X^{m_2} H Z_{\pm \alpha_2} X^{m_1} H Z_{\alpha
  1}|+\rangle$, where $m_1$ and $m_2$ are the outputs of the
measurements on the first and second qubits, respectively. Observe
that feedforward can be used to choose the sign of $\pm \alpha_2$ so
that $Z_{\pm \alpha_2} X^{m_1} = X^{m_1} Z_{\alpha_2}$.  We also have
$H X^{m_1} = Z^{m_1} H$, and thus the output may be rewritten as
$X^{m_2} Z^{m_1} H Z_{\alpha_2} H Z_{\alpha 1}|+\rangle$, which, up to
the known Pauli matrix $X^{m_2} Z^{m_1}$, is identical to the output
of the conventional single-qubit quantum
circuit~(\ref{eq:basic-one-qubit-circuit}).

This example generalizes easily to larger single-qubit circuits
containing gates of the form $HZ_\alpha$.  The general proof strategy
is: (1) rewrite the cluster-state computation in terms of an
equivalent quantum circuit; (2) reinterpret the quantum circuit as a
sequence of circuits of the form~(\ref{eq:transport}); (3) in the
resulting expression for the output state, commute operators of the
form $X^m$ all the way to the left, using feedforward to choose signs
on the terms of the form $Z_{\pm \alpha}$ to ensure that after
commutation they are of the form $Z_\alpha$.  The result is a state
which, up to a known Pauli matrix, is equivalent to the output of the
single-qubit quantum circuit.

These ideas generalize also to multi-qubit quantum circuits.  For
example, the circuit:
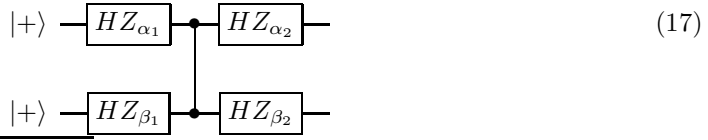
\begin{equation} 
\Qcircuit @C=1em @R=1.8em @!R {
 \lstick{|+\rangle} & \gate{HZ_{\alpha_1}} & \ctrl{1}     & 
\gate{HZ_{\alpha_2}} & \qw \\
 \lstick{|+\rangle} & \gate{HZ_{\beta_1}}  & \control \qw & 
\gate{HZ_{\beta_2}} & \qw 
}
\end{equation}
can be simulated using the cluster-state
computation~(\ref{eq:first-cluster-figure}).  The proof of this
equivalence follows exactly the same lines as in the single-qubit
case, and is only notationally more complicated.  We omit the details,
and suggest the interested reader fill them in.

Summing up, we have shown how the cluster-state model of computation
can be used to efficiently simulate any quantum circuit whose inputs
are all $|+\rangle$ states, and whose gates are either controlled-{\sc
  phase} gates, or gates of the form $HZ_\alpha$.  This set of
resources is universal for quantum computation, and thus the
cluster-state model is capable of efficiently simulating any quantum
circuit.  Conversely, it is straightforward to see that any
cluster-state computation may be efficiently simulated in the quantum
circuit model, and thus the two models are computationally equivalent.

\section{Properties of the cluster-state model}
\label{sec:misc}

We've given a basic description of the cluster-state model of quantum
computing.  In this section we describe two simple but fundamental
questions about cluster states, and present some progress on answering
these questions.  The results and methods presented are elementary,
but hopefully instructive, and strongly suggestive of a rich and
undiscovered structure in the theory of measurement-based quantum
computation.

Subsection~4.1 asks when a cluster state can arise as the ground state
of a quantum system, and shows that for typical graphs it is not
possible for the cluster state to be the ground state of a realistic
quantum system.  Subsection~4.2 addresses the question of what general
properties of a quantum state enable it to serve as a substrate for
quantum computation, in a manner similar to the cluster state.  We
show that the two-dimensional geometry of the cluster is important, by
demonstrating that a wide class of one-dimensional analogues of the
cluster-state model can be simulated efficiently on a classical
computer, and thus are unlikely to be useful for quantum information
processing.

\subsection{Cluster states and the ground states of many-body quantum systems}
\label{subsec:ground-states}

Given the significance of cluster states for quantum computation, it
is natural to ask whether or not cluster states may occur as
non-degenerate ground states of some naturally occurring class of
physical systems.  If this were so then cooling and measuring such a
system might offer a viable path to quantum computation.
Unfortunately, we show in this subsection that this is typically not
possible, provided we make a restriction that is usually physically
reasonable, namely, that the system has only two-body interactions.
The argument we give is based on~\cite{Haselgrove03b}, which studied
the conditions under which quantum error-correcting code states can
arise as ground states of Hamiltonian systems.

Our results are framed in terms of the stabilizer formalism introduced
by Gottesman~\cite{Gottesman97a} (see Chapter~10 of~\cite{Nielsen00a}
for a review).  Using the two-stage preparation procedure for the
cluster state associated to a graph $G$ it is easy to verify that a
set of generators for the stabilizer group of the cluster state is
given by the operators
\begin{eqnarray}
  S_v \equiv X_v \bigotimes_{v'} Z_{v'},
\end{eqnarray}
where there is one such stabilizer generator for each vertex, $v$,
$X_v$ and $Z_v$ represent Pauli operators acting on the qubit
associated to vertex $v$, and the tensor product is over all vertices
$v'$ neigbouring $v$.  Up to an unimportant global phase factor the
cluster is the unique quantum state such that $S_v |\psi\rangle =
|\psi\rangle$ for all vertices $v$ in the graph.  As an aside, we note
that this observation gives us a way of effectively preparing cluster
states using measurements alone (c.f.~\cite{Raussendorf03a}): (1)
measure all the stabilizer generators, obtaining a state which is a
simultaneous eigenstate of the $S_v$, with eigenvalues $\pm 1$; and
then (2) make use of the result~\cite{Gottesman97a,Nielsen00a} that
such a state is equal, up to local unitary operations, to the state
for which all the $S_v$ have eigenvalue $+1$, i.e., the cluster state.

Suppose $|C\rangle$ is a cluster state associated to some fixed graph,
$G$.  Can $|C\rangle$ arise as the non-degenerate ground state of some
two-body Hamiltonian, $H = \sum_{\sigma \tau} h_{\sigma \tau} \sigma
\otimes \tau$, where the sum is over distinct pairs of Pauli ($I, X,
Y, Z$) operators\footnote{We assume that interactions only occur
  between qubits whose corresponding vertices are connected; more
  general interaction topologies may be studied using similar
  techniques to those described here.}?  To answer this question,
observe that the state $(\sigma \otimes \tau) |C\rangle$ has a
stabilizer generated by the operators $(\sigma \otimes \tau) S_v
(\sigma \otimes \tau) = n^{\sigma,\tau}_v S_v$, where
$n^{\sigma,\tau}_v = \pm 1$.  We call the vector $n^{\sigma,\tau}$
whose entries are the $n^{\sigma,\tau}_v$ the \emph{syndrome vector}
corresponding to $\sigma \otimes \tau$.  Clearly, provided the
syndrome vectors $n^{\sigma,\tau}$ are all distinct then the states
$(\sigma \otimes \tau)|C\rangle$ are orthonormal.  When this occurs,
$H|C\rangle$ must contain terms orthonormal to $|C\rangle$, and thus
$|C\rangle$ cannot be an eigenstate of $H$.
Indeed,~\cite{Haselgrove03b} generalizes this argument, showing that
when this condition holds, the distance between $|C\rangle$ and the
energy eigenstates may be bounded below by a \emph{constant}
independent of anything except the Hilbert space dimension.
See~\cite{Haselgrove03b} for details.  When these properties hold we
say the cluster satisfies the \emph{unique syndrome condition}.

This argument can also be generalized in other ways.  Observe that if
$H$ is to be non-degenerate, then no qubit can be isolated, i.e.,
every qubit must interact with at least one other qubit through $H$.
Suppose the graph is such that we can find a qubit $v$ with the
following property: for all possible interaction terms $\sigma \otimes
\tau$ between $v$ and other qubits the syndrome vector
$n^{\sigma,\tau}$ is unique among \emph{all} the possible syndromes
$n^{\sigma',\tau'}$.  Once again, it is clear that under these
conditions $H|C\rangle$ must contain non-zero terms orthogonal to
$|C\rangle$, and thus $|C\rangle$ cannot be an eigenstate.  We say
that a vertex $v$ with these properties satisfies the \emph{unique
  syndrome condition}.  To establish that a cluster can not arise as a
ground state it therefore suffices to show that there is at least one
vertex satisfying the unique syndrome condition.

For which cluster geometries is there a vertex satisfying the unique
syndrome condition?  In a linear cluster it is not difficult to verify
that the unique syndrome condition does not hold for any vertex $v$,
and thus these techniques do not give any insight into whether or not
such clusters can arise as the ground state.  To see this, consider a
segment of three cluster qubits:
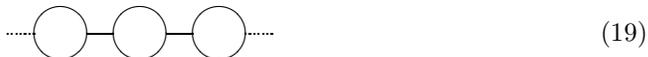
\begin{equation}
\Qcircuit[2em] @R=1em @C=1em { 
& \node{} \halflink{0}{-1} & \node{} \link{0}{-1} & \node{} \link{0}{-1}
 & \halflink{0}{-1}
}
\end{equation}
Using the stabilizer arguments above we see that $ZII$ and $IXZ$ give
rise to the same syndrome, and thus $ZII|C\rangle = IXZ|C\rangle$ (up
to a global phase), and so neither the second nor the third vertex can
satisfy the unique syndrome condition.  A similar argument holds for
other vertices in a linear cluster.

The situation changes if we move to two-dimensional clusters.  For
example, if we consider a toric cluster, then computation of all the
possible cases shows that for any vertex $v$ and for any interaction
terms $\sigma \otimes \tau$ between $v$ and other qubits the syndrome
vector $n^{\sigma,\tau}$ is unique among \emph{all} the possible
syndromes $n^{\sigma',\tau'}$.  It follows that the distance between
the cluster state and the non-degenerate ground state of a two-body
Hamiltonian whose couplings respect the cluster topology is bounded
below by a constant.

For rectangular lattices the situation changes because of the
boundary.  Consider the cluster:
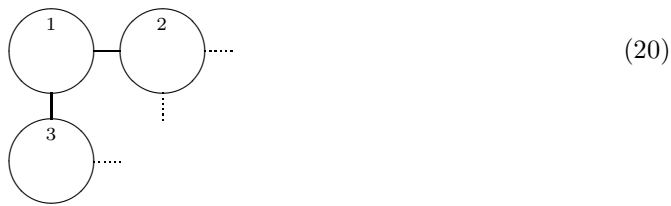
\begin{equation}
\Qcircuit[3.2em] @R=1em @C=1em {
\node[1]{} \link{1}{0} & \node[2]{} \link{0}{-1}         & \halflink{0}{-1} \\
\node[3]{}             & \halflink{0}{-1} \halflink{-1}{0} & 
}
\end{equation}
A computation shows $X_1 Z_2 I_3$ has the same syndrome as $I_1 I_2
Z_3$, and so our conditions cannot hold for all vertices.  However, if
we consider a vertex $v$ in the interior, then case enumeration shows
that for any interaction $\sigma \otimes \tau$ between $v$ and other
qubits the syndrome $n^{\sigma,\tau}$ is unique among \emph{all}
possible syndromes $n^{\sigma',\tau'}$.  Thus the unique syndrome
condition holds for this vertex, and so this cluster cannot arise as
the non-degenerate ground state of a two-body Hamiltonian whose
couplings respect the cluster topology.

Similar computations may be carried out for a wide variety of
clusters.  Consideration of some examples shows that, generically, the
clusters arising in simulations of non-trivial multi-qubit quantum
circuits cannot be non-degenerate ground states.  However, the question
of obtaining a general classification of which clusters satisfy these
conditions is still open.  More generally, it would be of great
interest to develop a general understanding of which clusters can and
cannot arise as ground states of physically reasonable Hamiltonians.
Note that when the use of ancilla qubits is allowed, the techniques
of~\cite{Kempe04a} may be used to obtain any cluster state as an
approximate non-degenerate ground state of a reasonable Hamiltonian.
However, these techniques have the disadvantage that the resulting gap
to the first excited state may be very small, and so the cluster state
may not be stable to thermal fluctuations.

\subsection{Linearly assembled quantum states can be efficiently simulated 
  on a classical computer}
\label{subsec:linear-states}

In this subsection we investigate what makes cluster states useful
substrates for quantum computation.  We show that spatial dimension
plays a role, proving that quantum states which can be linearly
prepared in one dimension can always be simulated efficiently on a
classical computer, and thus are not useful for quantum computation.

By a \emph{linearly preparable} quantum state we mean a state
$|\psi\rangle$ that can be prepared using a circuit of the form:
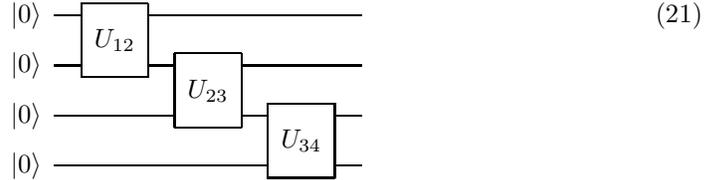
\begin{equation} 
\Qcircuit @C=1em @R=1em {
 & \lstick{|0\rangle} & \multigate{1}{U_{12}} & \qw                   &
\qw                   & \qw  \\
 & \lstick{|0\rangle} & \ghost{U_{12}}        & \multigate{1}{U_{23}} & 
\qw                   & \qw \\
 & \lstick{|0\rangle} & \qw                   & \ghost{U_{23}}        & 
\multigate{1}{U_{34}} & \qw  \\
 & \lstick{|0\rangle} & \qw                   & \qw                   & 
\ghost{U_{34}}        & \qw 
}
\end{equation}
i.e., a cascaded sequence of two-qubit quantum gates applied to some
product starting state.  It is easy to show that the two-dimensional
cluster states we have been using are not linearly preparable.

Suppose we perform a sequence of quantum measurements on a linearly
assembled quantum state, e.g.:
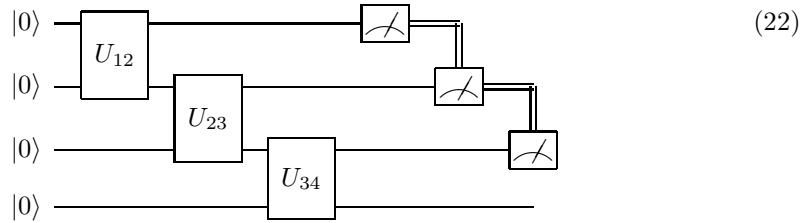
\begin{equation} 
\Qcircuit @C=1em @R=1em {
 & \lstick{|0\rangle} & \multigate{1}{U_{12}} & \qw                   &
\qw                   & \meter & \cw  \\
 & \lstick{|0\rangle} & \ghost{U_{12}}        & \multigate{1}{U_{23}} & 
\qw                   & \qw    & \meter \cwx & \cw \\
 & \lstick{|0\rangle} & \qw                   & \ghost{U_{23}}        & 
\multigate{1}{U_{34}} & \qw    & \qw         & \meter \cwx \\
 & \lstick{|0\rangle} & \qw                   & \qw                   & 
\ghost{U_{34}}        & \qw    & \qw         & \qw 
}
\end{equation}
where earlier measurement results may be used to control later
measurement bases.  Note that the measurement gates above may be in
any single-qubit basis, not just the computational basis.  This is
equivalent to the following circuit:
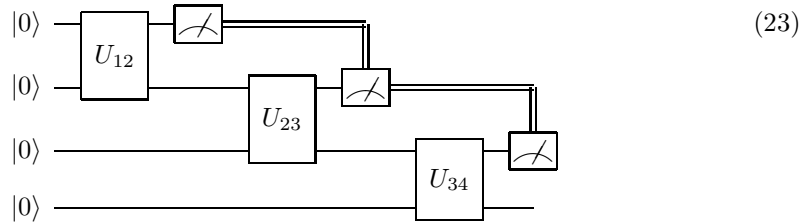
\begin{equation} 
\Qcircuit @C=1em @R=1em {
 & \lstick{|0\rangle} & \multigate{1}{U_{12}} & \meter & \cw                &
\cw \\
 & \lstick{|0\rangle} & \ghost{U_{12}}        & \qw & \multigate{1}{U_{23}} & 
\meter \cwx & \cw           & \cw \\
 & \lstick{|0\rangle} & \qw                   & \qw & \ghost{U_{23}}        & 
\qw & \multigate{1}{U_{34}} & \meter \cwx \\
 & \lstick{|0\rangle} & \qw                   & \qw & \qw                   & 
\qw & \ghost{U_{34}}        & \qw
}
\end{equation}
It is easy to use a classical computer to simulate such a circuit.
First, we consider the result of applying $U_{12}$ to $|00\rangle$,
and compute the probabilities for the measurement on qubit $1$.  We
then sample from this distribution to produce a posterior state for
qubit $2$, which is used as an input for $U_{23}$.  We then repeat
this cycle of computing probabilities, sampling, and computing
posterior states.  In order that the overall result be accurate, it
suffices to carry out computations to $O(\log(1/n))$ bits of
precision, and thus the entire computation can be carried out using
$O(n \log^c(1/n))$ operations, where $c$ is some constant arising from
the need to multiply floating point numbers.  Thus, efficient
classical simulation of such a computation can be performed.

What happens if we measure the qubits in some other order, e.g.,
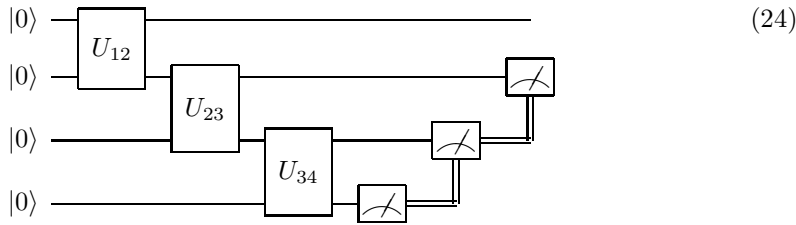
\begin{equation} 
\Qcircuit @C=1em @R=1em {
 & \lstick{|0\rangle} & \multigate{1}{U_{12}} & \qw                   &
\qw                   & \qw    & \qw            & \qw \\
 & \lstick{|0\rangle} & \ghost{U_{12}}        & \multigate{1}{U_{23}} & 
\qw                   & \qw    & \qw            & \meter \cwx[1] \\
 & \lstick{|0\rangle} & \qw                   & \ghost{U_{23}}        & 
\multigate{1}{U_{34}} & \qw    & \meter \cwx[1] & \cw  \\
 & \lstick{|0\rangle} & \qw                   & \qw                   & 
\ghost{U_{34}}        & \meter & \cw
}
\end{equation}

An elaboration of our earlier argument takes care of this case, too.
Define $\rho_j$ to be the state of the $j$th qubit after $U_{j-1 \,
  j}$ has been applied, but before $U_{j \, j+1}$ has been applied.
Using an argument similar to that earlier, it is not difficult to
determine $\rho_j$ for all $j$.

Suppose there are $n$ qubits involved; in our example, $n=4$.  When we
measure the $n$th qubit we induce trace-decreasing operations ${\cal
  E}_{n,0}$ and ${\cal E}_{n,1}$ on qubit $n-1$, corresponding to the
two possible measurement outcomes. These operations can be computed in
a compact form using standard methods (see, e.g.,~\cite{Nielsen00a}).
By computing the trace of these operations applied to $\rho_{n-1}$ we
can simulate the measurement statistics on the $n$th qubit.

To simulate the measurement on the $n-1$th qubit, we first compute the
two possible trace-decreasing operations ${\cal E}_{n-1,0}$ and ${\cal
  E}_{n-1,1}$ describing the change in state of qubit $n-2$,
conditional on the two possible measurement outcomes on qubit $n-1$.
We now simulate the measurement statistics on the $n-1$th qubit by
computing the trace of these operations applied to $\rho_{n-2}$,
renormalizing according to the probability of the measurement outcome
obtained on the $n$th qubit.  Iterating this procedure, we can
simulate all the different measurement outcomes using a classical
computer.  In order to be accurate, computations are carried out to
$O(n)$ bits of accuracy, and so this procedure introduces a time
overhead which is polynomial in $n$, and so is efficient.

This efficient classical simulation procedure can easily be
generalized to a sequence of measurements performed in any order on
the qubits, including adaptive orderings.  Furthermore, it is
straightforward to extend these results to some simple variants on the
linear topology, including rings, and a small (e.g., constant) number
of interlinking rings.

It remains an interesting open problem to determine what class of
states can be used as a substrate for quantum computation.  The
examples in this subsection indicate that the geometry underlying the
preparation procedure plays a key role.  It would be interesting to
develop a better understanding of what that role is, and what, if
anything, that tells us about the physical properties responsible for
the power of quantum computing.

\section{Conclusion}
\label{sec:conc}

We have described the basic ideas behind the quantum circuit and
cluster-state models of quantum computation, including the proof of
the remarkable fact that measurements on a cluster state can be used
to simulate the unitary operations at the heart of the quantum circuit
model.  We have also illustrated the cluster state model with some
simple observations about two questions of fundamental interest: (a)
can the cluster state arise as the ground state of a naturally
occurring Hamiltonian; and (b) what properties make the cluster state
a useful substrate for quantum computation?  Our results illustrate
the rich structure of the cluster-state model, and emphasize the
importance of further investigations of this structure.

\section*{Acknowledgments}

Thanks to Sean Barrett, Hans Briegel, Andrew Doherty, Jens Eisert,
Daniel Gottesman, Henry Haselgrove, Charles Hill, David Mermin, and
Tim Ralph for helpful comments, and to Steve Flammia and Bryan Eastin
for adding cluster state support to their \LaTeX ``Qcircuit'' package,
which was used to produce the figures in this paper.  The results on
efficient classical simulation of linearly prepared states were
obtained jointly with Andrew Doherty, and I thank Andrew for
permission to include them here.

%\bibliographystyle{plain}
%\bibliography{../../mybib}

\end{document}